\documentclass[graybox,vecphys]{svmult}

\usepackage{makeidx}         
\usepackage{graphicx}        
\usepackage{multicol}        
\usepackage[bottom]{footmisc}

\usepackage{url} 

\makeindex             

\begin{document}

\newcommand{\jcmindex}[2]{\index{{\bf\large #1}!#2}}
\newcommand{\jcmindext}[3]{\index{{\bf\large #1}!#2!#3}}

\title*{Nonequilibrium phenomena in nonlinear lattices:
from slow relaxation to anomalous transport}
\titlerunning{Nonequilibrium nonlinear lattices}
\author{
Stefano Iubini
\and
Stefano Lepri
\and 
Roberto Livi
\and
Antonio Politi
\and 
Paolo Politi
}

\institute{
 Stefano Iubini
\at
 Dipartimento di Fisica e Astronomia, Universit\`a di Padova, via F. Marzolo 8 I-35131, Padova, Italy
\email{stefano.iubini@unipd.it}
\and
 Stefano Lepri
\at
Consiglio Nazionale
delle Ricerche, Istituto dei Sistemi Complessi, 
via Madonna del Piano 10, I-50019 Sesto Fiorentino, Italy
\email{stefano.lepri@isc.cnr.it}
\at
Istituto Nazionale di Fisica Nucleare, Sezione di Firenze, 
via G.  Sansone 1 I-50019, Sesto Fiorentino, Italy
\and
Roberto Livi 
\at
Dipartimento di Fisica e Astronomia, Universit\`a di Firenze, 
via G. Sansone 1 I-50019, Sesto Fiorentino, Italy
\email{roberto.livi@unifi.it}
\at
Istituto Nazionale di Fisica Nucleare, Sezione di Firenze, 
via G.  Sansone 1 I-50019, Sesto Fiorentino, Italy
\at
Consiglio Nazionale
delle Ricerche, Istituto dei Sistemi Complessi, 
via Madonna del Piano 10, I-50019 Sesto Fiorentino, Italy
\and
Antonio Politi
\at
Institute for Pure and Applied Mathematics \& SUPA University of Aberdeen, Aberdeen AB24 3UE, Scotland, United Kingdom
\email{a.politi@abdn.ac.uk}
\and
Paolo Politi
\at
Consiglio Nazionale
delle Ricerche, Istituto dei Sistemi Complessi,
via Madonna del Piano 10, I-50019 Sesto Fiorentino, Italy
\email{paolo.politi@isc.cnr.it}
\at
Istituto Nazionale di Fisica Nucleare, Sezione di Firenze, 
via G.  Sansone 1 I-50019, Sesto Fiorentino, Italy
}
\maketitle
\abstract
{
This Chapter contains an overview of the effects of nonlinear interactions
in selected problems of non-equilibrium statistical mechanics.
Most of the emphasis is put on open setups, where energy is exchanged
with the environment.
With reference to a few models of classical coupled anharmonic oscillators,
we review anomalous but general properties such as extremely slow relaxation
processes, or non-Fourier heat transport.
}

\section{Introduction}

The title of this Chapter contains two negations, \textit{nonequilibrium}
and \textit{nonlinearity}, which signal a double source of difficulties.
First, at variance with equilibrium statistical mechanics, 
there is no general approach to describe the evolution of  a generic
system far from equilibrium. Second, nonlinear forces
notoriously have to be handled with care. 

From the fundamental point of view, nonlinear interactions are 
essential for the theoretical foundations of irreversible processes:
a derivation of phenomenological relations (like for instance 
Fourier's law) from
microscopic dynamics is indeed one of the challenges of mathematical physics.
On the other hand, understanding the role of 
nonlinearity, low-dimensionality, long-range interactions, disorder etc.
may help developing innovative ideas for nanoscale
thermal management with possible future applications
like controlling the heat fluxes in small 
devices built on molecular junctions, carbon nanotubes, polymers and 
nano-structured materials \cite{Lepri2016}.

This Chapter aims at illustrating
the combined effect of nonlinear interactions on relaxation and transport: 
since this is an exceedingly
vast topic, we focus mainly on selected specific issues.
In particular, we wish to review through some examples
(mostly relying on numerical simulations) of how relaxation and transport are 
affected by nonlinear interactions in systems of classical nonlinear oscillators.
This class of model represents
a large variety of different physical problems like atomic vibrations
in crystals and molecules or field modes in optics or acoustics.

The Chapter is organized as follows. For concreteness, we discuss 
mostly one-dimensional arrays of classical oscillators, that
are reviewed in Section~\ref{sec:models}. For later purposes, their 
equilibrium thermodynamics is recalled in Section~\ref{sec:equilibrium}.
Section~\ref{sec:relax} deals with the typical times-scales of 
relaxation to a steady state in the presence of a dissipation applied to
the boundaries and discuss how nonlinear localization can significantly affect the process.
This is the most detailed section, since the topic is still open and we have preferred to
add some recent details for the sake of clarity.
In Section~\ref{sec:transport} we address the issue of nonequilibrium steady states
for chains in contact with different thermal reservoirs. We recall 
there how nonlinear interactions affect fluctuations of conserved quantities
and conspire to yield energy superdiffusion.
Finally, in Section \ref{sec:end} we outline possible future 
developments.

\section{Classical coupled nonlinear oscillators: basic models}
\label{sec:models}

A vast number of micro- and mesoscopic models have been introduced and studied to 
understand nonequilibrium dynamics. Many of them involve some 
form of stochasticity \cite{Livi2017}. Here we concentrate on open Hamiltonian models 
described by unidimensional arrays of $N$ classical nonlinear oscillators.
Two families of models are reviewed: (i) separable systems characterized by
kinetic and potential energy; (ii) non-separable ones such as the
the Discrete Nonlinear Schr\"odinger (DNLS) equation.

The first class is generally characterized by the Hamiltonian
\begin{equation}
{H} = \sum_{n=1}^N \left[{p_n^2\over 2m} + U(q_n) +
V(q_{n+1}-q_{n})\right] \quad ,
\label{optical}
\end{equation}
where $q_n$ and $p_n$ denote position and momentum of the point--like particles; $m$ is their mass
while the potential $V(x)$ accounts for the nearest-neighbour interactions between consecutive particles,
and, finally, the on-site potential $U(q_n)$ accounts for the possible interaction with an
external environment (e.g., a substrate).

The corresponding evolution equations are
\begin{equation} 
m{\ddot q}_n = -U'(q_n) - F(r_{n}) + F(r_{n-1})  \quad , \quad n=1,\ldots,N \, ,
\label{eqmot} 
\end{equation}  
where $r_n=q_{n+1}-q_n$, $F(x)=- V'(x)$, and the prime denotes a derivative with respect to the argument. If $q_n$ denotes a longitudinal position, then
$L= \sum_{n=1}^{N-1} r_n$
represents the total length of the chain, which, in the case of fixed boundary conditions (b.c.), is a constant of motion.
Conversely, if the particles are confined in a simulation ``box"
of length $L$ with periodic b.c., we have  $q_{n+N} \;=\; q_{n} \,+\, L $.
Alternatively, one can adopt a lattice interpretation whereby
the (discrete) position is $x_n = an$ (where $a$ is the lattice spacing), while $q_n$
is a transversal displacement. Thus, the chain length is obviously equal to $Na$.

For isolated systems, the Hamiltonian (\ref{optical}) is a constant of motion.
If the pinning potential $U$ is constant, the total momentum 
$P= \sum_{n=1}^N p_n$
is conserved, as well. 
Since we are interested in heat transport, one can set $P = 0$ 
(i.e., we assume to work in the center--of--mass reference frame) without
loss of generality. 
As a result, the relevant state variables of microcanonical equilibrium are the 
specific energy (i.e., the energy per particle) $h=H/N$ and the 
elongation $\ell=L/N$ (i.e., the inverse of the particle density).

An important subclass is the one in which $V$ is quadratic, which can be regarded as a discretization
of the Klein-Gordon field: relevant examples are the Frenkel-Kontorova \cite{Gillan85,Hu1998}
and ``$\phi^4$'' models \cite{Aoki00,Kevrekidis2019} which, in suitable units, correspond to 
$U(y) = 1-\cos(y)$ and $U(y) = y^2/2+y^4/4$, respectively.              
Another toy model that has been studied in some detail 
is the ding-a-ling system~\cite{Casati84}, where $U$ is quadratic and the nearest-neighbor
interactions are replaced by elastic collisions.

\subsection{The Fermi-Pasta-Ulam-Tsingou chain}

In this context, the most paradigmatic example is the Fermi--Pasta--Ulam-Tsingou (FPUT) model,
with $U(q_n)=0$ and
\begin{equation}
V(r_n)\;=\; \frac{k_2}{2}\,(r_n-a)^2+ \frac{k_3}{3}\,(r_n-a)^3 + 
\frac{k_4}{4}\, (r_n-a)^4 \quad ,
\label{fpu}
\end{equation} 
introduced in a widely acknowledged seminal work~\cite{Fermi1955} in nonlinear dynamics.
It is well known that the initial goal of the study was to demonstrate that a generic
nonlinear interaction should eventually drive an isolated mechanical system with many degrees of freedom, towards an equilbrium state characterised by energy equipartition among
normal modes. Actually, in the following decades the related 
problem of steady state transport was also considered \cite{Payton67,Nakazawa1970,Kaburaki93}.

Following the notation of the original work \cite{Fermi1955}, the coupling terms
$k_3$ and $k_4$ are denoted by $\alpha$ and $\beta$, respectively;
historically this model is sometimes referred to as the ``FPU-$\alpha\beta$'' model.
In the absence of the cubic nonlinearity ($k_3=0$), the system is referred to as ``FPU-$\beta$'' model. 
Notice that upon introducing the displacement $u_n = q_n - na$ from the equilibrium position,
$r_n$ can be rewritten as $u_{n+1}-u_n+a$, so that the lattice spacing $a$ disappears from the equations.

\subsection{The Discrete nonlinear Schr\"odinger equation}

The Discrete Nonlinear Schr\"odinger (DNLS) equation has been 
widely investigated in various domains of physics as a prototype 
model for the propagation of nonlinear excitations \cite{Eilbeck1985,Eilbeck2003,Kevrekidis}. 
Originally, it was proposed to describe electronic transport in 
biomolecules \cite{Scott2003} and later for nonlinear wave propagation in 
photonic or phononic crystals \cite{Kosevich02,Hennig99} as well as in ultra-cold atom gases 
in optical lattices \cite{Franzosi2011}.

The system (in its dimensionless form) is described by the Hamiltonian 
\begin{equation}
\label{Hz}
H= \sum_{n=1}^N \left( |z_n|^4+z_n^*z_{n+1}+z_nz_{n+1}^* \right) \quad 
\end{equation}
where the complex variables $z_n$ and $-iz_n^*$ ($n=1,\ldots,N$) are canonical variables.
The Hamilton equations $\dot{z}_n=-\partial{H}/\partial{(iz_n^*)}$ are written as
\begin{equation}
\label{eqmotdnls}
i \dot{z}_n=-2|z_n|^2z_n -z_{n-1}- z_{n+1} \; .
\end{equation}
Sometimes it is convenient to decompose $z_n$ into real and imaginary components:
$z_n = (p_n + i q_n)/\sqrt{2}$; this way $q_n$ and $p_n$ are standard conjugate canonical variables. 
Besides the Hamiltonian, the system admits a second constant of motion, namely the total norm
$A = \sum_{n=1}^N |z_n|^2 $ which, depending on the physical context, can be interpreted as the gas particle number, 
optical power, etc.
At variance with its continuum counterpart, the DNLS is non-integrable: it typically displays a chaotic dynamics.

A peculiarity of this model is the existence of  localized solutions, the so-called {\it discrete breathers} 
(DB) \cite{Kevrekidis,Flach2008}, characterized
by a large amplitude on a single site, $|z_n|^2 \gg s^2$, where $s^2$ is the amplitude of the surrounding background.
In the limit of $s\ll 1$, when perturbative calculations can be carried out, the long term stability has been
discussed in full detail \cite{Johansson2004}. We later on show that breather stability is an important issue also in physical setups,
where the background is fully chaotic.


\subsection{The coupled rotors model} 

Another interesting system is the coupled rotors chain described by the equations of motion
\begin{equation}\label{2}
\dot q_n = p_n\,,\quad \dot p_n = \sin(q_{n+1} - q_n)- \sin(q_{n} - q_{n-1})\,.
\end{equation}
This model is sometimes referred to as the Hamiltonian version of the XY spin chain. 
It is a sort of intermediate model between standard oscillator chains (notice that here
the ``position'' $q_n$ is an angle) and the DNLS equation, once we think of the variable
$z_n$ as composed of amplitude and phase. In fact, it can be shown that in some limit the DNLS reduces to
a rotor chain \cite{Iubini2013a}. 

\section{Equilibrium}
\label{sec:equilibrium}

Equilibrium thermodynamics of the above models can be calculated by standard
means. For the Hamiltonian (\ref{optical}) without pinning ($U=0$) this 
can be accomplished straightforwardly by computing the partition function in the isobaric 
ensemble \cite{Spohn2014}. For models like Klein-Gordon and DNLS lattices, the computation
requires the use of transfer integral methods \cite{Rasmussen2000}.

The case of the DNLS is of particular interest:
a thermodynamic equilibrium state is in fact specified by two intensive 
parameters, the mass density $a=A/N\geq 0$ and the energy density $h=H/N$ or equivalently 
by the conjugate variables $\mu$ (chemical potential) and $\beta$ (inverse temperature). 
The equilibrium phase-diagram in the $(a,h)$ plane \cite{Rasmussen2000}
is bounded from below by the 
$(T=0)$ ground--state line $h=a^2-2a$ corresponding to a uniform state with constant amplitude and constant phase--differences
$z_n=\sqrt{a}e^{i(\mu t+\pi n)}$, with $\mu=2(a-1)$. 
States below this curve are not physically accessible. 

The positive--temperature region lies 
between the ground--state line and  the infinite-temperature $(\beta=0)$ line, given by $h=2a^2$.
In this limit, the grand--canonical equilibrium distribution becomes proportional to $\exp{(\beta\mu A)}$,
where the finite (negative) product $\beta\mu$ implies a diverging chemical potential. 
Equilibrium states at infinite temperature are therefore characterized by an exponential distribution of the amplitudes,
$P(|z_n|^2)=a^{-1}e^{-|z_n|^2/a}$ and random phases. Finally, states above the $\beta=0$ line
belong to the so--called negative--temperature region~\cite{Rasmussen2000,Iubini2013}.
From a thermodynamic point of view, the presence of states at absolute negative temperature
is a consequence of the entropy being a decreasing function of the internal energy. 
Very recently it has been found that this region is characterized by inequivalence of
statistical ensembles \cite{Gradenigo2019}. 
While just above the $\beta =0$ line the microcanonical partition function
can be computed explicitly by large-deviation techniques, the grandcanonical partition function is undefined, due to the
presence of a branch-cut singularity in the complex $\beta$-plane. Moreover, the microcanonical ensemble
predicts the presence of a first-order phase transition from a thermalized phase, below the $\beta =0$ line,
to a condensed phase, above the $\beta =0$ line. This situation represents a typical scenario of
broken ergodicity in the negative-temperature phase, induced by a condensation phenomenon, due to the spontaneous formation
and coalescence of DB.

Since the  DNLS Hamiltonian (\ref{Hz}) is not separable, one cannot determine temperature and chemical potential 
via the standard molecular-dynamics tools.
It is necessary to make use of the microcanonical definition provided in~\cite{Rugh1997}. The general expressions are 
nonlocal and rather involved; we refer to \cite{Iubini2012} for details and the related bibliography. Alternatively, one can 
determine the relations $a(T,\mu)$, $h(T,\mu)$ numerically, by putting the system in interaction with an external reservoir 
that imposes $T$ and $\mu$ and by measuring the corresponding equilibrium densities.
\footnote{
The actual implementation of a reservoir for the DNLS is less straightforward
than for usual oscillator models \cite{LLP03}. 
Two main strategies have been proposed: the first 
is a Monte-Carlo dynamics \cite{Iubini2012} whereby the reservoir performs random perturbations $\delta z_1$
of, say, the state variable $z_1$ that are accepted
or rejected according to a grand-canonical Metropolis cost-function 
$\exp{[-\beta(\Delta H-\mu \Delta A)]}$,
where $\Delta H$ and $\Delta A$ are respectively the variations of energy and mass produced by 
$\delta z_1$. Between successive interactions with the environment the dynamics is
Hamiltonian and can be integrated by symplectic algorithms \cite{Yoshida1990}.
Another approach is based on a Langevin dynamics with a dissipation 
designed in such a way that equilibrium corresponds to the grand-canonical measure \cite{Iubini2013}.
}

\section{Relaxation}
\label{sec:relax}

In this section we review and discuss relaxation dynamics, namely how the equilibrium
steady state is reached starting from nonequilibrium initial conditions. Generally speaking
one may distinguish between two cases: 
\begin{itemize}
\item[(i)]
relaxation to thermal equilibrium (energy equipartition) from
a particular initial state in the \textit{isolated} (microcanonical) setup; 
\item[(ii)]
~evolution towards
a steady state in an \textit{open} setup, whereby the system is allowed to 
exchange energy, momentum etc. with the environment, composed of one (or more) reservoirs. 
\end{itemize}

A typical example of case (i) is the numerical experiment discussed in the original FPUT paper,
a problem deeply related to the validity of the ergodic hypothesis.
In the following, we will not dwell further on the FPUT problem: 
the interested reader can look at some recent literature \cite{Benettin2011,Benettin2013,Onorato2015}. 
Here, we focus more on boundary-induced relaxation phenomena with a particular emphasis given to the DNLS, 
for the existence of a negative-temperature region. 

\subsection{Localization by boundary cooling}

A first numerical evidence of slow-relaxation induced by the spontaneous emergence of localized inhomogeneities
has been provided by Tsironis and Aubry~\cite{Tsironis1996}, who discussed the chain dynamics in the presence 
of a nonlinear pinning potential. The system, initially prepared in a thermalized state 
at some given temperature $T$, was put in contact with a cold (zero temperature) heat bath,
by adding a damping term on a few boundary particles. 
The chain eventually converges to a quasi-stationary state,
where a residual amount of energy is kept under the form of a
few isolated DBs. It has been later argued that the energy relaxation obeys a stretched exponential law in 
time~\cite{Piazza2001,Piazza2003}.

The rotor model is yet another interesting example where boundary dissipation leads to a slow 
relaxation. In Ref.~\cite{Eleftheriou2005} it was found that for long enough times the 
energy decreases very slowly, according to a typical stretched-exponential law
\[
E(t)= E(0) \exp (-(t/\tau)^\sigma ) 
\]
with $\sigma<1$ (typically $\sigma\approx 0.5$) and $\tau$ being some characteristic timescale. 
The occurrence of slow dynamics has been associated with a progressive destruction of localized excitations
(the so-called rotobreathers)  and energy release at the boundaries. At very long times a residual 
quasi-stationary state is again observed: a finite fraction of the initial
energy is stored into a single rotobreather and remains constant over the rest of the simulation.

Recently, some mathematical insight on the origin of such slow process has been proposed:
it has been argued that the dissipation rate may become arbitrarily small in certain physical regimes due to the decoupling of non-resonant terms, as it happens 
in KAM problems
\cite{Cuneo2016,Cuneo2017}.

A renewed interest in the problem of boundary cooling was provided by the proposal
of implementing it as a technique to localize Bose-Einstein condensates in optical lattices \cite{Livi2006}. Referring to the DNLS model, stationary and traveling localized states were generated by removing atoms at the optical-lattice ends. Regimes of stretched-exponential decay for the number of atoms trapped in the lattice were clearly identified by numerical simulations. Further studies showed that the dynamics
of dissipated energy exhibits a characteristic avalanche behavior \cite{Ng2009}.

\subsection{Dynamical freezing of relaxation to equilibrium}

In this Section we discuss the relaxation process of a localized excitation (DB)
in the DNLS model (see Eqs.~(\ref{Hz}-\ref{eqmotdnls})).
For positive temperatures, DBs have a negligible probability to arise
at equilibrium and yet, as commented above, there are ways to grow them (e.g., by
boundary dissipation). Therefore, it is important to understand their relaxation process.

Simulations have been performed by superposing a large-mass DB at the origin, $n=0$, to an
otherwise equilibrium configuration for the background (i.e. $-N\le n\le N$).
The temperature $T=10$ and the chemical potential $\mu=-6.4$ are imposed by connecting the chain
boundaries to suitable Langevin baths. 

\begin{figure}
\includegraphics[width=\textwidth]{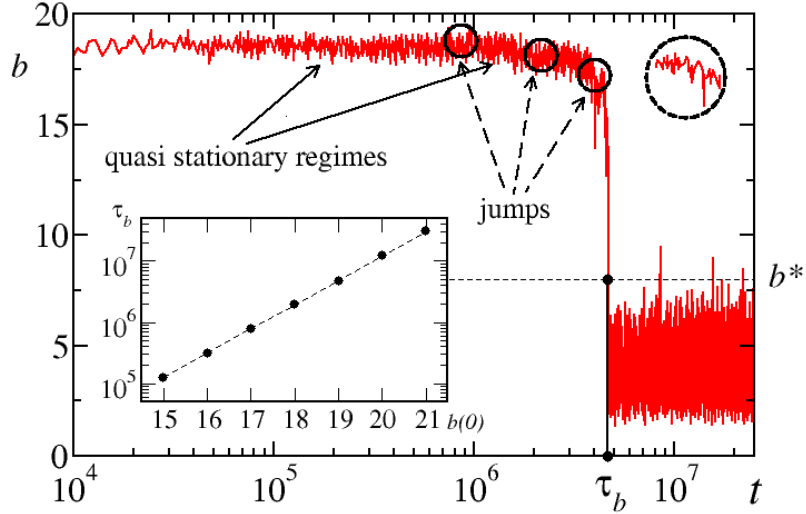}
\caption{The relaxation process of a breather through the time
dependence of its mass, $b(t)$. The abrupt character of the process allows to define
a threshold value $b^*$ and to determine the relaxation time $\tau_b$
as the shortest time satisfying the condition $b(\tau_b)=b^*$.
In the top right circle we zoom in a jump of the mass.
The inset shows the exponential increase of $\tau_b$ with the initial mass
of the breather: the dashed line is a fit, giving $\tau_b \approx e^{\alpha b(0)}$, with
$\alpha=0.91\pm 0.01$.
The breather is initially at the centre of a chain of $N=31$ sites.
}
\label{tau_b_dnls}
\end{figure}

The main results are reported in Fig.~\ref{tau_b_dnls}, where we can see
the time dependence of the breather mass $b = |z_0|^2$ in a typical simulation
(notice the logarithmic time scale).
The decay process is fairly abrupt and we can identify
quasi-stationary regimes, separated by jumps.
Because of the abrupt decay of $b(t)$ we can define the relaxation time
$\tau_b$ by setting a threshold $b^*$ (dashed horizontal time).
In the inset, we show how $\tau_b$ depends on the initial mass of the breather,
namely exponentially.
In the rest of this Section we will discuss the meaning of this numerical finding
and some related results.

The rapid increase of $\tau_b$ with $b(0)$ suggests that the larger the mass of the breather,
the weaker  the relaxation mechanism. This is not surprising, 
because the natural frequency of a breather of mass $b$ is 
$\omega=2b$, so the coupling term between the breather and the nearest neighbors
becomes negligible on the typical time scale of the background (which is of order one).
However, a rotational frequency proportional to the breather mass 
cannot justify by itself an exponentially slow decay process.
Before trying any theoretical explanation, it is necessary to gain more insight on
the relaxation process.

From Fig.~\ref{tau_b_dnls} we can see that the {\it laminar} regime preceding 
the final breather disruption is approximately quasi-stationary, with no specific
drift (except for the final part, where the mass clearly tends to decrease).
It is, therefore, tempting to characterize this regime in terms of a suitable diffusion 
coefficient\footnote{Bounded fluctuations would be possible only in the presence
of an attractor, but this is a Hamiltonian system.}.
The next question is the identification of an optimal variable to characterize the
hypothetical diffusion process.
The mass $b(t)$ is too noisy to extract reliable estimates.
A principal component analysis (PCA) performed on the triplet of variables $z_{-1},z_0,z_1$
suggests to use $\tilde Q \simeq E_b^{1/4}$~\cite{Iubini2019}, where
$E_b = |z_0|^4 + \frac{1}{2}[z_0^*(z_1 +z_{-1}) + \mbox{ c.c.}]$ is the breather energy (see Eq.~(\ref{Hz})).
Accordingly we define
$D_{\tilde Q} =\langle [ \tilde Q^2(t+\tau) -\tilde Q^2(t)]\rangle /\tau$.
The evaluation of $D_{\tilde Q}$ for breathers of increasing height
(see Fig.~4 of Ref.~\cite{Iubini2019})
shows that the diffusion coefficient decreases exponentially with $b(0)$,
in agreement with the direct evidence of an almost frozen dynamics.
On the basis of fluctuation-dissipation considerations, we should
also expect a drift $v \approx D_{\tilde Q} /T$ and, accordingly,
an exponential increase of the decay time.
However, while the exponential increase is confirmed by the simulations,
we do not see a clear evidence of a (downward) drift.

\begin{figure}
\includegraphics[width=\textwidth]{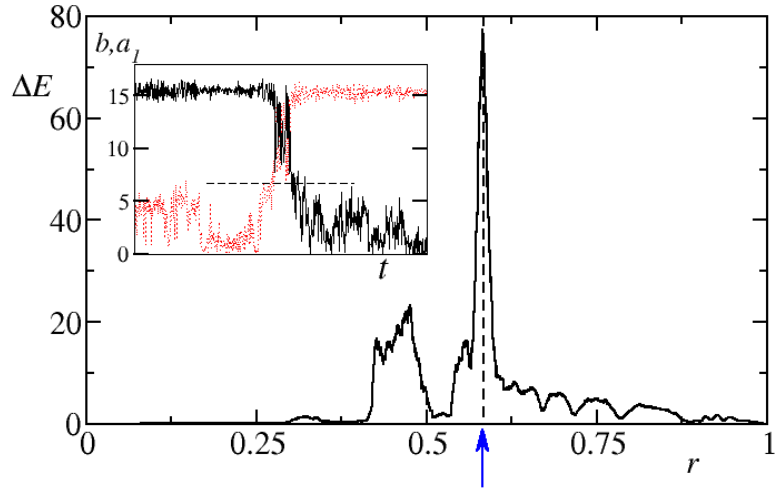}
\caption{The energy lost or gained on average by the breather after a fixed amount
of time, as a function of the initial mass of the neighbouring site, $a_1(0)$
(more precisely, of the ratio $r=a_1(0)/b(0)$).
The sharp peak corresponds to the condition 
$\sqrt{a_1(0)} = \sqrt{b(0)} - \sqrt{2}$,
which allows the formation of a symmetric bound state (dimer) between the
breather and its neighbour.
In the inset we plot the time dependence of the mass in these two sites
along with such threshold value (horizontal dashed line).}
\label{dimer_peak}
\end{figure}

Fig.~\ref{tau_b_dnls} rather suggests that the relaxation
is accompanied and perhaps caused by sporadic jumps. 
This statement is supported by Fig.~\ref{dimer_peak},
where we plot the average transfer of energy between the breather
and the background in a given interval of time, as a function of
the initial mass of the neighbouring site, $a_1(0)$ (suitably rescaled).
There is an evident, narrow peak accompanied by some pinnacles.
The peak appears at a value very close to the analytical
threshold for the existence of symmetric bound states (dimers) between
two neighbouring sites, $\sqrt{a_1^*(0)} = \sqrt{b(0)} -\sqrt{2}$.

The very emergence of such bound states is attested by the inset of the same
figure, where we plot the time dependence of the mass of the breather
$b(t)$ along with the mass $a_1(t)$ of a neighboring site.
When it happens that $a_1 > a_1^*$ (see the dashed line)
the two sites are strongly coupled together and rotate with the same
frequency. This bound state eventually dissolves, with the 
net result that the ``post-dimer" breather has lost some energy
with respect to the ``pre-dimer" one.

Analytical, non-rigorous considerations allow evaluating
the typical time scale of such phenomena as the expectation
time for a background fluctuation in one of the two neighbouring sites
to become larger than the threshold $a_1^*$. In fact, 
from the high-$T$ equilibrium distribution $P(\theta)$ to have a 
mass $\theta$~\cite{Rasmussen2000}, we can approximately estimate the
breather lifetime $\tau_b \approx 1/P(a_1^*)$, which has the
asymptotic expression $\tau_b \approx \exp(\beta b^2)$
for a diverging breather mass $b$.
This expression implies a superexponential growth of the relaxation time 
with the mass, suggesting that the asymptotic mechanism for breather
decay might not be dimer formation ($\tau_b$
seems to increase exponentially), but current simulation data
do not allow to exclude it either. 

\begin{figure}
\includegraphics[width=\textwidth]{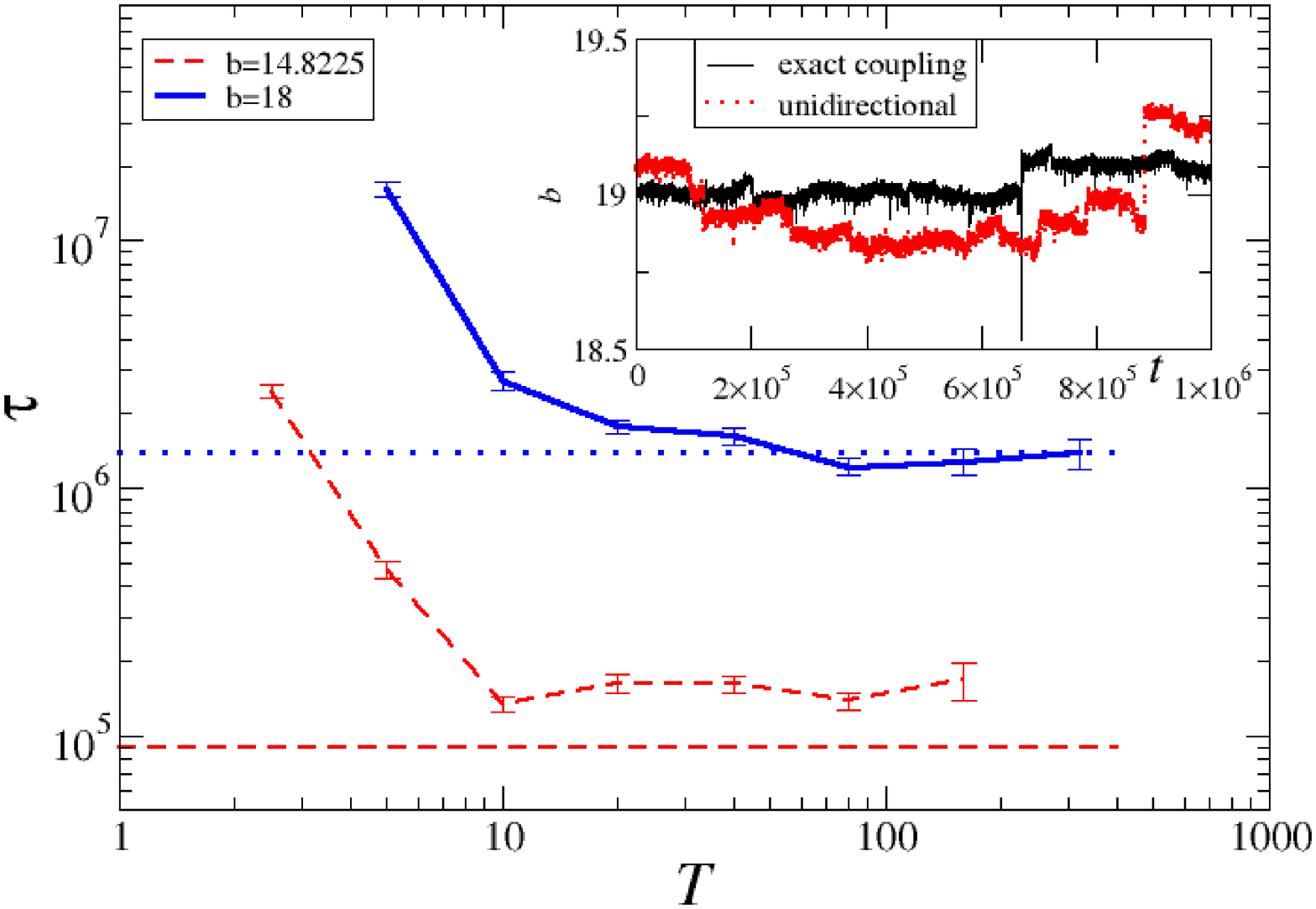}
\caption{The relaxation time of a breather as a function of the temperature $T$,
for two different initial masses. The horizontal lines represent the analytical,
asymptotic values as deduced by an approximate approach (see the main text) 
where the breather feels the background but not the other way around.
In the inset we compare the time evolution of the mass of the same initial
breather, with exact and unidirectional coupling.}
\label{tau_T_dnls}
\end{figure}

As a matter of fact, the additional peaks appearing in Fig.~\ref{dimer_peak}
suggest the existence of other mechanisms and 
the special setup we are going to discuss allows one to conclude that
relaxation occurs even if dimer formation is suppressed.
We direct reader's attention to the inset of Fig.~\ref{tau_T_dnls}, where
we consider the relaxation process of the very same breather
in different conditions. The full line is the standard DNLS 
model, i.e. a curve fully similar to that plotted in Fig.~\ref{tau_b_dnls}.
The dotted line has been determined with a unidirectional coupling between breather
and background: the former feels the latter but not the other way around.
In practice the breather is coupled to a background whose evolution
is independent of the breather itself.
The result is a qualitatively similar evolution of the breather mass.

Since unidirectional coupling 
makes impossible the rising of a symmetric bound state,
in this case dimers cannot play any role in DB relaxation.
On the other hand, resonance mechanisms would not be affected by the
unidirectional character of the coupling and therefore they would persist,
but the approximate expression for the relaxation time would still
give a superexponential growth.

Unidirectional coupling is useful also to discuss the temperature dependence of 
the relaxation time, $\tau_b(T)$.
The curves plotted in Fig.~\ref{tau_T_dnls}
refer to different initial masses of the breather and the central result
we focus on is the asymptotic behavior of $\tau_b(T)$, which seems to reach a constant.
In other words, the relaxation time of a finite breather is expected to be finite
even at infinite temperature. The horizontal lines have been determined analytically
with the same criterion discussed here above, taking the limit $T\to\infty$
but also allowing $|\mu|=T/a$ to diverge so that the density of mass, $a$, 
is kept constant (simulations have been performed within such setup).

So far, we have mentioned two potentially interesting processes:
(i) the diffusion occurring during the laminar (quasi-stationary) regime;
(ii) the jumps, both visible in Fig.~\ref{tau_b_dnls}.
The diffusion coefficient turns out to decrease exponentially with $b$ 
and the probability of dimer or resonances formation 
(related to the jumps) also decreases exponentially or even superexponentially.
It is evident that in the presence of several relaxation channels,
a frozen dynamics may appear if and only if all mechanisms are exponentially slow.

However, it is not clear at all {\it why} dynamics is almost frozen. 
A former paper by some of the present authors~\cite{Iubini2019} - a first attempt in this direction -
suggests that $\tilde Q$ (the quantity derived from the PCA analysis and used
to derive the diffusion coefficient) is the approximate expression of an
Adiabatic Invariant (AI), which might be broken by jumps.
We have tried to determine AI perturbatively, by computing higher orders, but the
attempt has not been successful.

We wish to stress that understanding the relaxation process of a breather
at positive temperature is a useful if not necessary step to later understand the negative 
$T$ phase from a dynamical point of view.

\subsection{Role of negative temperatures}

In the previous section we have seen that relaxation to equilibrium may be very slow
in the DNLS, in the positive temperature region of the $(a,h)$ plane.
As mentioned in Section \ref{sec:equilibrium}, there exists a second region
($h>2a^2$) where the absolute temperature is expected to be negative (on the basis
of microcanonical arguments).
Relaxation phenomena in this region turn out to be a rather controversial issue, not yet
fully settled.

Entropic arguments \cite{Rumpf2004,Rumpf2008} suggest that all the excess energy, which cannot be stored in a 
homogeneous background for $h>2a^2$, should eventually concentrate into a single breather.
In fact, this is precisely what happens in a simplified, purely stochastic version of 
the DNLS, where it has been shown that multiple breathers progressively merge through 
a non conventonal coarsening process~\cite{Iubini2014,Iubini2017a}.

\begin{figure}
\includegraphics[width=0.8\textwidth,clip]{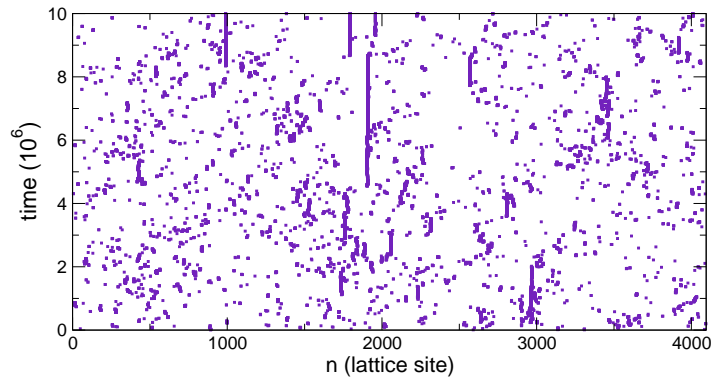}
\caption{Evolution of the local amplitude for DNLS in a negative temperature
state, the dots correspond to points where $|z_n|^2 > 10$.
Microcanonical simulation with $a=1$ and $h=2.4$, $N=4192$.}
\label{fig:dbspacetime}
\end{figure}

On the other hand, coalescence has not been observed in molecular dynamics simulations slightly above
the $\beta=0$ line (e.g. for $a=1$ and $h=2.4$). On the contrary, it looks like a sort of 
stationary regime sets in, characterized by a small breather density, where DBs 
spontaneously form and then die, after some typical lifetime \cite{Iubini2013}, see 
Fig.~\ref{fig:dbspacetime}. This is due to the presence of a finite interaction (hopping) energy: the background can store excess energy in the phase differences of neighbouring sites and this implies that breathers can spontaneously nucleate. 
These findings have been recently confirmed by more extensive simulations \cite{Mithun2018} 
which suggest the existence of a finite region in the $(a,h)$-plane, where the dynamics
covers a subregion of the available phase space, doing so in an ``ergodic manner" (i.e. no 
coarsening).

Further, independent evidence of a relatively stable negative-temperature regime comes
from the nonequilibrium simulations performed in \cite{Iubini2017}, where a 
DNLS chain was put in contact
on one side with a positive temperature heat bath, while on the other, with a pure
dissipator. Depending on the temperature value of the first heat bath, an extended portion of
the chain, settles in a regime characterized by a position-dependent negative temperature,
a flux of mass and energy, without being accompanied by the onset of breathers.

Finally, recent statistical-mechanics calculations 
\cite{Gradenigo2019} suggest that slightly above 
the critical $\beta=0$ line, strong finite-size effects are to be expected, which 
might affect the interpretation of the numerical simulations.

\section{Transport}
\label{sec:transport}

Let us now turn our attention to nonequilibrium steady states that emerge, at long enough times, when 
the system is in contact with two (or more) heat reservoirs operating at different temperatures.
Generally speaking, several methods, based on both deterministic
and stochastic (Langevin or Monte-Carlo) algorithms, have been proposed \cite{LLP03}.
A complementary approach is based on linear-response theory, which amounts to computing
the equilibrium correlation function of currents.  In principle, this task
can be accomplished in any equilibrium ensemble, the 
microcanonical one being the most natural choice.

\subsection{Anomalous energy transport}

The main results, emerging from a long series of works, can be summarized
as follows. Models of the form (\ref{eqmot}) with $U(q)=0$ typically display  
\textit{anomalous} transport and relaxation features.
Said differently, 
Fourier's law \textit{does not hold}: the kinetics of energy carriers is so
correlated that they are able to propagate \textit{faster} than in 
the standard (diffusive) case. 
We refer the reader to existing review papers \cite{LLP03,DHARREV,Basile08,Lepri2016}
for a more comprehensive description. Here we just mention how this anomalous  behaviour manifests itself
in the simulations.
\begin{itemize}
\item The finite-size heat conductivity $\kappa(L)$  diverges  in
the limit of a large system size $L\to \infty$ as , 
$\kappa(L) \;\propto\; L^\gamma$  \cite{LLP97}, 
i.e. the heat transport coefficient is ill-defined in the thermodynamic limit.

\item The equilibrium correlation function of the total
energy current $J$ displays a nonintegrable long-time tail 
$\langle J(t)J(0)\rangle \propto t^{-(1-\delta)}$, with $0\le\delta < 1$  
\cite{Lepri98a,L00}. Accordingly, 
the Green-Kubo formula yields an infinite value of the conductivity.

\item Energy perturbations propagate superdiffusively \cite{Denisov03,Cipriani05}: 
a local perturbation of the energy spreads, while  its variance broadens in time as
$\sigma^2\propto t^{\beta}$,
with $\beta > 1$.

\item Temperature profiles in the nonequilibrium steady states 
are nonlinear, even for vanishing applied temperature 
gradients. Typically they are the solution of a \textit{fractional heat equation}
\cite{Lepri2011b,Kundu2019}.

\end{itemize}

There is a large body of numerical evidence that the above features
occur generically in 1D and 2D, whenever the conservation of energy, 
momentum and length holds. This is related to the existence of
long-wavelength (Goldstone) modes  (an acoustic phonon branch in the linear
spectrum of (\ref{eqmot}) with $U=0$) that are very weakly damped.
Indeed, it is sufficient to add external (e.g. substrate) forces, to make
all anomalies disappear and restore Fourier's law.

\subsection{Universality and the Kardar-Parisi-Zhang equation}

The nonlinear fluctuating hydrodynamics approach is able to justify and
predict several universal features of anomalous transport in anharmonic chains
\cite{Spohn2014,VanBeijeren2012}. The main entities are 
the random fields describing deviations of the conserved quantities 
with respect to their stationary values. The role of fluctuations
is taken into account by renormalization group or some kind of 
self-consistent theory. 

The main theoretical insight is the intimate relation between the 
anharmonic chain and 
one of the most important equations in nonequilibrium statistical 
physics, the celebrated Kardar-Parisi-Zhang (KPZ) equation, 
originally introduced in the (seemingly unrelated) context of 
surface growth \cite{Barabasi1995}. 
The KPZ equation for the stochastic field $h(x,t)$ in one spatial dimension reads
\begin{equation}
\frac{\partial h}{\partial t}=
\nu\frac{\partial^2 h}{\partial x^2}+
\frac{\kappa} 2\left(\frac{\partial h}{\partial x}\right)^2 
 +  \eta.
\label{kpzh}
\end{equation}
where $\eta(x,t)$ represents a Gaussian white noise with 
$\langle\eta(x,t)\eta(x',t')\rangle$=$2D\delta(x-x')\delta(t-t')$
and $\nu,\kappa,D$ are the relevant parameters. 
It has been shown \cite{Spohn2014,VanBeijeren2012} that large-scale 
dynamical properties of anharmonic chains are in the 
same dynamical universality class as Eq.(\ref{kpzh}).
Loosely speaking, we can represent the displacement field as the superposition 
of counter-propagating plane waves, modulated by an envelope that is ruled, 
at large scales, by Eq.~(\ref{kpzh}).
As a consequence, correlations of observables display in the hydrodynamic
limit \textit{anomalous dynamical scaling}. For instance, the dynamical 
structure factor $S(k,\omega)$ of the particle displacement shows 
for $k\to 0$ two sharp peaks at $\omega = \pm\omega_{\rm max}(k)$
that correspond to the propagation of sound modes 
and for $\omega\approx\pm \omega_{\rm max}$ behave as
\begin{equation}\label{scaling32}
S({k},\omega)\sim f_{\rm KPZ}\left(\frac{\omega\pm\omega_{\rm max}}{\lambda_s k^{3/2}}\right).
\end{equation}
Remarkably, the scaling function $f_{\rm KPZ}$ is universal and known exactly, while
$\lambda_s$ is a model-dependent parameter.
The main point is that the \textit{dynamical exponent} $z=3/2$ is different from $z=2$ expected 
for a standard diffusive process.

Most of the predictions have been successfully tested for several models. 
For a chain of coupled anharmonic oscillators with three conserved quantities like 
the FPUT chains, such theoretical predictions have been successfully 
compared with the numerics \cite{Das2014a,DiCintio2015}.
Other positive tests have been reported in Ref.~\cite{Mendl2013}.
One further prediction is that 
the FPUT-$\beta$ model should belong to a different (non-KPZ) universality class,
as previously suggested by the numerics \cite{Lepri03,Lee-Dadswell2015}.

\subsection{Coupled transport}

As known from irreversible thermodynamics, when there are more conserved quantities, 
the corresponding currents can be coupled: in the linear response 
regime, transport is described by Onsager coefficients.
The best-known example is that of thermoelectricity, whereby useful electric work 
can be extracted in the presence of temperature gradients \cite{Mejia2001,Casati2009,Benenti2014}.  

In the present context, the simplest example is 
the one-dimensional rotor model, Eq. (\ref{2}),  
that admits two conserved quantities (energy and
angular momentum), two associated currents, 
and only one relevant thermodynamic parameter, the temperature $T$.
In this case \cite{Iubini2014} one can easily introduce the interaction 
with two reservoirs by fixing the 
average angular momenta $\omega_0,\omega_1$ and kinetic temperatures
$T_0,T_1$ at the chain ends. This can be obtained by adding the Langevin term 
$\gamma(\omega_0 - \dot q_1) + \sqrt{2\gamma T_0}\,\xi$ 
to the equation of motion (\ref{2}) of the leftmost rotor: $\gamma$ defines the coupling strength with the bath
and $\xi$ is a Gaussian white random noise 
with zero mean and unit variance. An analogous term, with
$\omega_1$ and $T_1$ replacing $\omega_0$ and $T_0$, is added
to the equation of motion of the rightmost rotor. 

\begin{figure}
\includegraphics[width=0.8\textwidth]{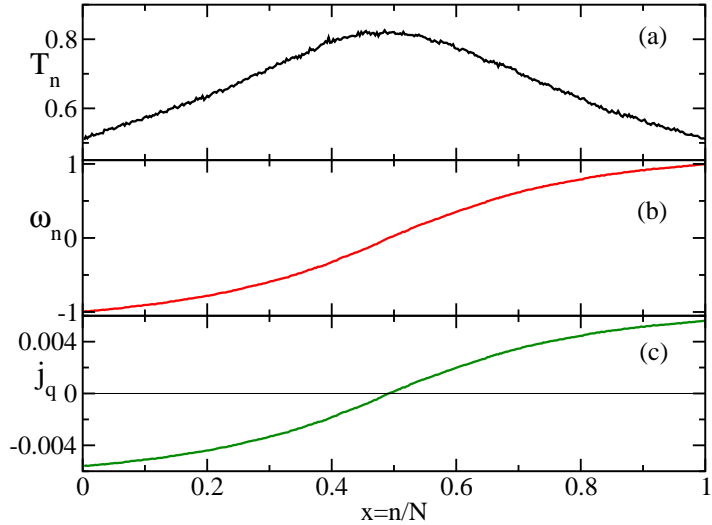}
\caption{Simulation of the rotor chain with $N=400$ particles, in contact at its 
boundaries with two
heat baths  at
temperature $T_0 = T_1 =  0.5$ and in the presence of  torques $\omega_0=-1$ and
$\omega_1=1$: (a) temperature
profile; (b) frequency (chemical potential) profile; (c)  local heat flux. }
\label{fig:profile}
\end{figure}

To illustrate the peculiarities of this setup, Fig.~\ref{fig:profile} reports frequency and temperature profiles \cite{Iubini2016}
in a case where only  an angular momentum gradient is applied, i.e. $T_1=T_0$ and
$\omega_0 = -1$ and $\omega_1=1$. 
The temperature profile $T_n$ is non-monotonic \cite{Iacobucci2011} as 
a consequence of the coupling with the momentum flux $j_p$ imposed by 
the torque at the boundaries, although, in the end, the energy flux $j_h$
vanishes for symmetry reasons.
By recalling that $j_h = j_q + \omega j_p$ we see that the heat flux
$j_q = - \omega j_p$ varies along the chain being everywhere proportional
to the frequency, so that it is negative in the left part and positive in
the right side (this is again consistent with symmetry considerations).
Thus heat is generated in the central hotter region,
where the temperature is higher and transported towards the two
edges. The total energy flux is however everywhere zero as the heat flux
is compensated by an opposite coherent flux due to momentum transfer.
Physically, the temperature bump can be interpreted as a sort of Joule effect: the transport
of momentum involves dissipation, which in turn contributes to increasing
the temperature, analogously to what happens when an electric wire is crossed by a flux
of charges.

Similar effects are studied in \cite{Iubini2012} for the DNLS case, where the dependence
of the Onsager matrix on temperature and chemical potential is considered.
In this case the cross-coupling term (the equivalent of the Seebeck coefficient in
the language of thermo-electricity) may change sign, leading to temperature- and mass-profiles 
with opposite slopes.

\subsection{Integrable models and their perturbations}

The above results are mostly obtained in a strongly nonlinear regime or more 
generally far from any integrable limit. For the FPUT model, this means working 
with high enough energies/temperatures to avoid all the difficulties induced by quasi-integrability and the associated slow relaxation to equilibrium.

Integrable systems constitute \textit{per se} a relevant case. In the framework
of the present work the most important example is certainly the celebrated 
Toda chain, namely model (\ref{optical}) with $U=0$ and 
\[
V(x) = e^{-x} + x -1 
\] 
As intuitively expected, heat transport is ballistic due to its 
integrability and the associated solitonic solution \cite{Toda79}. Mathematically,
this is expressed by saying that there is a non-vanishing \textit{Drude weight},
namely a zero-frequency component of the energy current power spectra \cite{Zotos02,Shastry2010}.
A lower bound of the Drude weight can be estimated making use of Mazur inequality \cite{Mazur1969}
in terms of correlations between the currents themselves and the conserved quantities 
(see \cite{Zotos02,Shastry2010} for the Toda case).
However, the 
idea of solitons transporting energy as independent particles is
somehow too simplistic. It has been recognized \cite{Theod1999} that solitons 
experience a stochastic sequence of spatial shifts as they move through the lattice
interacting with other excitations without momentum exchange \cite{Theod1999}.
At variance with the harmonic chain,
which is also integrable, but whose proper modes are non-interacting phonons,
the Toda chain, as proposed in \cite{Spohn2018}, is an interacting integrable system.
In particular, it is characterized by what has been termed a non--dissipative diffusion
mechanism \cite{Theod1999}.
In fact, the calculation of the transport coefficients by the Green-Kubo formula indicates the
presence of a finite Onsager coefficient, which corresponds to a diffusive process on top of
the dominant ballistic one \cite{Shastry2010,Kundu2016,DiCintio2018}.

A natural question concerns the behavior when a generic perturbation is applied
to an otherwise integrable system. For instance, adding a quadratic pinning potential 
$U(q)=q^2/2 $ to the Toda chain is expected to restore standard diffusive transport, but 
numerical simulations show that long-range correlations are preserved over
relatively long scales \cite{DiCintio2018,Dhar2019}. Moreover, weak perturbations that 
conserve momentum (and are thus expected to display anomalous transport in the KPZ class)
display instead significant differences \cite{Iacobucci2010} 
and even diffusive transport over the accessible simulation ranges \cite{Chen2014}. 
Altogether a full unified picture of the problem is still lacking.

\section{Overview and open problems}
\label{sec:end}

In spite of the sensible progress that has been made over the last decades, the study of
nonequilibrium processes in nonlinear systems remains a fascinating and challenging domain
of research. On a methodological ground, it concerns mainly the scientific communities of 
mathematics  and theoretical physics, but it is of primary interest also for optics, materials science and
soft matter, just to mention a few among the related fields in experimental and  applied research.
Going through the reading of this Chapter, one can easily realize that most achievements have been
possible thanks to a fruitful combination of analytic approaches and numerical simulations and this
can be reasonably expected to hold also in the future. For what concerns open problems,
it is worth mentioning a few of them, that have already attracted some interest. The first one is
the study of nonlinear models with long-range interactions. The main motivation stems from the
observation that for this class of models the equivalence between statistical ensembles may not
hold. This is expected to yield interesting consequences also for nonequilibrium phenomena.
In fact, long-range  systems are known to exhibit further peculiar properties, like long-living metastable 
states, anomalous energy diffusion, lack of thermalization when interacting with a single temperature 
reservoir, propagation of perturbations with infinite velocity, etc. (for a general review see \cite{Campa2014}).
The problem of heat transport in the long-range version of the rotor and FPUT chains has been 
recently tackled in a series of papers \cite{avila2015length, Bagchi2017, Olivares2016, Iubini2018,DiCintio2019}. 
When  the interaction is genuinely long-range, i.e. the long-range exponent $\alpha$ is smaller than 1,
the heat transport process is dominated by parallel transport: a flat temperature profile sets in,
simply because each oscillator, independently, takes a temperature value which is the average of
those applied by the thermal baths. For $\alpha$ larger than 1, the long-range rotor model reproduces 
standard diffusion, i.e. normal heat conductivity, as in the short-range version. Conversely, the FPUT chain
is characterized by an anomalous scaling exponent $\gamma(\alpha)$, which seems to recover the
value of the short-range case only for  large values of $\alpha$. Anyway, a better understanding of
the transport problem demands further refined investigations, that should take into
account also a comparison with the unusual relaxation process to equilibrium characterizing long-range systems.

Let us conclude by mentioning one further open problem, which is related to 
energy localization induced by nonlinearity. In fact, localization processes may emerge in nonlinear
systems even in the absence of disorder. The typical example is the spontaneous formation of
breathers in the DNLS problem already discussed in Section \ref{sec:relax}. In the negative 
temperature region the phenomenon of condensation of DBs can be read as a process of ergodicity-breaking,
because, in the microcanonical setup, energy equipartition is inhibited by its localization. It is well
known that, when condensation phenomena are present, statistical ensemble equivalence is not granted \cite{Ruelle1999}.
This is the case  of the DNLS Hamiltonian in the negative temperature region, where the only available
statistical ensemble is the microcanonical one \cite{Gradenigo2019}. It is worth recalling that both total energy and mass are
conserved quantities in the DNLS Hamiltonian: the breaking of ergodicity  in the negative temperature phase
indicates that a standard thermalization process for both quantities is suppressed.  The analogy with
the problem of the Eigenvalue Thermalization Hypothesis advanced for genuine quantum integrable systems 
(see the review paper \cite{Vidmar2016})
suggests that ensemble equivalence should be reconsidered also in this context 
to properly formulate a thermalization hypothesis or, alternatively, the many-body localization phenomenon
invoked in quantum integrable systems.

\bibliographystyle{unsrt}
\bibliography{heat,levy,books,diodo}

\begin{thebibliography}{10}

\bibitem{Lepri2016}
Stefano Lepri, editor.
\newblock {\em Thermal transport in low dimensions: from statistical physics to
  nanoscale heat transfer}, volume 921 of {\em Lect. Notes Phys}.
\newblock Springer-Verlag, Berlin Heidelberg, 2016.

\bibitem{Livi2017}
Roberto Livi and Paolo Politi.
\newblock {\em Nonequilibrium statistical physics: a modern perspective}.
\newblock Cambridge University Press, 2017.

\bibitem{Gillan85}
MJ~Gillan and RW~Holloway.
\newblock Transport in the {Frenkel-Kontorova} model 3: thermal-conductivity.
\newblock {\em J. Phys. C}, 18(30):5705--5720, 1985.

\bibitem{Hu1998}
Bambi Hu, Baowen Li, and Hong Zhao.
\newblock Heat conduction in one-dimensional chains.
\newblock {\em Phys. Rev. E}, 57(3):2992, 1998.

\bibitem{Aoki00}
K.~Aoki and D.~Kusnezov.
\newblock Bulk properties of anharmonic chains in strong thermal gradients:
  non-equilibrium $\phi^4$ theory.
\newblock {\em Phys. Lett. A}, 265(4):250, 2000.

\bibitem{Kevrekidis2019}
Panayotis~G Kevrekidis and Jesus Cuevas-Maraver.
\newblock {\em A Dynamical Perspective on the $\phi^4$ Model: Past, Present and
  Future}, volume~26.
\newblock Springer, 2019.

\bibitem{Casati84}
Giulio Casati, Joseph Ford, Franco Vivaldi, and William~M. Visscher.
\newblock One-dimensional classical many-body system having a normal thermal
  conductivity.
\newblock {\em Phys. Rev. Lett.}, 52(21):1861--1864, 1984.

\bibitem{Fermi1955}
Enrico Fermi, J~Pasta, and S~Ulam.
\newblock Studies of nonlinear problems.
\newblock {\em Los Alamos Report LA-1940}, page 978, 1955.

\bibitem{Payton67}
DN~Payton, M~Rich, and WM~Visscher.
\newblock Lattice thermal conductivity in disordered harmonic and anharmonic
  crystal models.
\newblock {\em Phys. Rev.}, 160(3):706--\&, 1967.

\bibitem{Nakazawa1970}
Hiroshi Nakazawa.
\newblock On the lattice thermal conduction.
\newblock {\em Progress of Theoretical Physics Supplement}, 45:231--262, 1970.

\bibitem{Kaburaki93}
H~Kaburaki and M~Machida.
\newblock Thermal-conductivity in one-dimensional lattices of
  {Fermi-Pasta-Ulam} type.
\newblock {\em Phys. Lett. A}, 181(1):85--90, SEP 27 1993.

\bibitem{Eilbeck1985}
J.~C. Eilbeck, P.~S. Lomdahl, and A.~C. Scott.
\newblock {The discrete self-trapping equation}.
\newblock {\em Physica D}, 16:318–338, 1985.

\bibitem{Eilbeck2003}
J.~C. Eilbeck and M.~Johansson.
\newblock The discrete nonlinear {Schroedinger} equation-20 years on.
\newblock In L.~Vazquez, R.~S. MacKay, and M.~P. Zorzano, editors, {\em
  Conference on Localization and Energy Transfer in Nonlinear Systems},
  page~44. World Scientific, Singapore, 2003.

\bibitem{Kevrekidis}
Panayotis~G. Kevrekidis.
\newblock {\em The Discrete Nonlinear Schrödinger Equation}.
\newblock Springer Verlag, Berlin, 2009.

\bibitem{Scott2003}
A.~Scott.
\newblock {\em Nonlinear science. Emergence and dynamics of coherent
  structures.}
\newblock Oxford University Press, Oxford, 2003.

\bibitem{Kosevich02}
A.~M. Kosevich and M.~A. Mamalui.
\newblock Linear and nonlinear vibrations and waves in optical or acoustic
  superlattices (photonic or phonon crystals).
\newblock {\em J. Exp. Theor. Phys.}, 95(4):777, 2002.

\bibitem{Hennig99}
D.~Hennig and G.P. Tsironis.
\newblock Wave transmission in nonlinear lattices.
\newblock {\em Phys. Rep.}, 307(5-6):333–432, 1999.

\bibitem{Franzosi2011}
R.~Franzosi, R.~Livi, G.L. Oppo, and A.~Politi.
\newblock {Discrete breathers in Bose–Einstein condensates}.
\newblock {\em Nonlinearity}, 24:R89, 2011.

\bibitem{Flach2008}
Sergej Flach and Andrey~V Gorbach.
\newblock Discrete breathers -- advances in theory and applications.
\newblock {\em Phys. Rep.}, 467(1):1--116, 2008.

\bibitem{Johansson2004}
Magnus Johansson and Kim Rasmussen.
\newblock Statistical mechanics of general discrete nonlinear{ Schroedinger
  models: Localization transition and its relevance for Klein-Gordon} lattices.
\newblock {\em Physical Review E}, 70(6):066610, 2004.

\bibitem{Iubini2013a}
S~Iubini, S~Lepri, R~Livi, and A~Politi.
\newblock Off-equilibrium {Langevin dynamics of the discrete nonlinear
  Schroedinger} chain.
\newblock {\em J. Stat. Mech: Theory Exp.}, (08):P08017, 2013.

\bibitem{Spohn2014}
Herbert Spohn.
\newblock Nonlinear fluctuating hydrodynamics for anharmonic chains.
\newblock {\em J. Stat. Phys.}, 154(5):1191--1227, 2014.

\bibitem{Rasmussen2000}
KØ Rasmussen, T.~Cretegny, Panayotis~G. Kevrekidis, and N.~Grønbech-Jensen.
\newblock Statistical mechanics of a discrete nonlinear system.
\newblock {\em Phys. Rev. Lett.}, 84(17):3740–3743, 2000.

\bibitem{Iubini2013}
S~Iubini, R~Franzosi, R~Livi, GL~Oppo, and A~Politi.
\newblock Discrete breathers and negative-temperature states.
\newblock {\em New J. Phys.}, 15(2):023032, 2013.

\bibitem{Gradenigo2019}
Giacomo Gradenigo, Stefano Iubini, Roberto Livi, and Satya~N. Majumdar.
\newblock Localization in the discrete non-linear {Schroedinger} equation:
  mechanism of a first-order transition in the microcanonical ensemble, 2019.
\newblock arXiv:1910.07461.

\bibitem{Rugh1997}
Hans~Henrik Rugh.
\newblock Dynamical approach to temperature.
\newblock {\em Phys. Rev. Lett.}, 78(5):772, 1997.

\bibitem{Iubini2012}
S.~Iubini, S.~Lepri, and A.~Politi.
\newblock Nonequilibrium discrete nonlinear {Schroedinger equation}.
\newblock {\em Phys. Rev. E}, 86(1):011108, 2012.

\bibitem{LLP03}
S~Lepri, R~Livi, and A~Politi.
\newblock Thermal conduction in classical low-dimensional lattices.
\newblock {\em Phys. Rep.}, 377:1, 2003.

\bibitem{Yoshida1990}
Haruo Yoshida.
\newblock Construction of higher order symplectic integrators.
\newblock {\em Phys. Lett. A}, 150(5-7):262--268, 1990.

\bibitem{Benettin2011}
G.~Benettin and A.~Ponno.
\newblock Time-scales to equipartition in the {Fermi--Pasta--Ulam} problem:
  Finite-size effects and thermodynamic limit.
\newblock {\em Journal of Statistical Physics}, 144(4):793, Aug 2011.

\bibitem{Benettin2013}
G.~Benettin, H.~Christodoulidi, and A.~Ponno.
\newblock The {Fermi-Pasta-Ulam} problem and its underlying integrable
  dynamics.
\newblock {\em Journal of Statistical Physics}, 152(2):195--212, Jul 2013.

\bibitem{Onorato2015}
Miguel Onorato, Lara Vozella, Davide Proment, and Yuri~V. Lvov.
\newblock Route to thermalization in the $\alpha$-{Fermi-Pasta-Ulam} system.
\newblock {\em Proceedings of the National Academy of Sciences},
  112(14):4208--4213, 2015.

\bibitem{Tsironis1996}
GP~Tsironis and S~Aubry.
\newblock Slow relaxation phenomena induced by breathers in nonlinear lattices.
\newblock {\em Phys. Rev. Lett.}, 77(26):5225, 1996.

\bibitem{Piazza2001}
F~Piazza, S~Lepri, and R~Livi.
\newblock Slow energy relaxation and localization in 1d lattices.
\newblock {\em J. Phys. A: Math. Gen.}, 34(46):9803, 2001.

\bibitem{Piazza2003}
Francesco Piazza, Stefano Lepri, and Roberto Livi.
\newblock Cooling nonlinear lattices toward energy localization.
\newblock {\em Chaos: An Interdisciplinary Journal of Nonlinear Science},
  13(2):637--645, 2003.

\bibitem{Eleftheriou2005}
Maria Eleftheriou, Stefano Lepri, Roberto Livi, and Francesco Piazza.
\newblock Stretched-exponential relaxation in arrays of coupled rotators.
\newblock {\em Physica D: Nonlinear Phenomena}, 204(3):230--239, 2005.

\bibitem{Cuneo2016}
Noe Cuneo and J-P Eckmann.
\newblock Non-equilibrium steady states for chains of four rotors.
\newblock {\em Communications in Mathematical Physics}, 345(1):185--221, 2016.

\bibitem{Cuneo2017}
Noe Cuneo, Jean-Pierre Eckmann, and C~Eugene Wayne.
\newblock Energy dissipation in hamiltonian chains of rotators.
\newblock {\em Nonlinearity}, 30(11):R81--R117, oct 2017.

\bibitem{Livi2006}
Roberto Livi, Roberto Franzosi, and Gian-Luca Oppo.
\newblock Self-localization of {Bose-Einstein} condensates in optical lattices
  via boundary dissipation.
\newblock {\em Phys. Rev. Lett.}, 97:60401, 2006.

\bibitem{Ng2009}
GS~Ng, Holger Hennig, Ragnar Fleischmann, Tsampikos Kottos, and Theo Geisel.
\newblock Avalanches of {Bose--Einstein} condensates in leaking optical
  lattices.
\newblock {\em New journal of physics}, 11(7):073045, 2009.

\bibitem{Iubini2019}
Stefano Iubini, Liviu Chirondojan, Gian-Luca Oppo, Antonio Politi, and Paolo
  Politi.
\newblock Dynamical freezing of relaxation to equilibrium.
\newblock {\em Phys. Rev. Lett.}, 122:084102, Mar 2019.

\bibitem{Rumpf2004}
Benno Rumpf.
\newblock Simple statistical explanation for the localization of energy in
  nonlinear lattices with two conserved quantities.
\newblock {\em Phys. Rev. E}, 69(1):016618, 2004.

\bibitem{Rumpf2008}
Benno Rumpf.
\newblock Transition behavior of the discrete nonlinear {Schroedinger
  equation}.
\newblock {\em Phys. Rev. E}, 77(3):036606, 2008.

\bibitem{Iubini2014}
Stefano Iubini, Stefano Lepri, Roberto Livi, and Antonio Politi.
\newblock Boundary-induced instabilities in coupled oscillators.
\newblock {\em Phys. Rev. Lett.}, 112:134101, 2014.

\bibitem{Iubini2017a}
Stefano Iubini, Antonio Politi, and Paolo Politi.
\newblock Relaxation and coarsening of weakly-interacting breathers in a
  simplified {DNLS} chain.
\newblock {\em Journal of Statistical Mechanics: Theory and Experiment},
  2017(7):073201, jul 2017.

\bibitem{Mithun2018}
Thudiyangal Mithun, Yagmur Kati, Carlo Danieli, and Sergej Flach.
\newblock Weakly nonergodic dynamics in the {Gross-Pitaevskii} lattice.
\newblock {\em Physical review letters}, 120(18):184101, 2018.

\bibitem{Iubini2017}
Stefano Iubini, Stefano Lepri, Roberto Livi, Gian-Luca Oppo, and Antonio
  Politi.
\newblock A chain, a bath, a sink, and a wall.
\newblock {\em Entropy}, 19(9), 2017.

\bibitem{DHARREV}
Abhishek Dhar.
\newblock Heat transport in low-dimensional systems.
\newblock {\em Adv. Phys.}, 57:457--537, 2008.

\bibitem{Basile08}
G.~Basile, L.~Delfini, S.~Lepri, R.~Livi, S.~Olla, and A.~Politi.
\newblock Anomalous transport and relaxation in classical one-dimensional
  models.
\newblock {\em Eur. Phys J.-Special Topics}, 151:85--93, 2007.

\bibitem{LLP97}
S~Lepri, R~Livi, and A~Politi.
\newblock Heat conduction in chains of nonlinear oscillators.
\newblock {\em Phys. Rev. Lett.}, 78(10):1896--1899, MAR 10 1997.

\bibitem{Lepri98a}
S~Lepri, R~Livi, and A~Politi.
\newblock On the anomalous thermal conductivity of one-dimensional lattices.
\newblock {\em Europhys. Lett.}, 43(3):271--276, AUG 1 1998.

\bibitem{L00}
S~Lepri.
\newblock Memory effects and heat transport in one-dimensional insulators.
\newblock {\em Eur. Phys J. B}, 18(3):441--446, DEC 2000.

\bibitem{Denisov03}
S.~Denisov, J.~Klafter, and M.~Urbakh.
\newblock Dynamical heat channels.
\newblock {\em Phys. Rev. Lett.}, 91(19):194301, 2003.

\bibitem{Cipriani05}
P.~Cipriani, S.~Denisov, and A.~Politi.
\newblock From anomalous energy diffusion to {Levy} walks and heat conductivity
  in one-dimensional systems.
\newblock {\em Phys. Rev. Lett.}, 94(24):244301, 2005.

\bibitem{Lepri2011b}
Stefano Lepri and Antonio Politi.
\newblock Density profiles in open superdiffusive systems.
\newblock {\em Phys. Rev. E}, 83(3):030107, 2011.

\bibitem{Kundu2019}
Aritra Kundu, Cedric Bernardin, Keji Saito, Anupam Kundu, and Abhishek Dhar.
\newblock Fractional equation description of an open anomalous heat conduction
  set-up.
\newblock {\em J. Stat. Mech: Theory Exp.}, 2019(1):013205, 2019.

\bibitem{VanBeijeren2012}
Henk van Beijeren.
\newblock Exact results for anomalous transport in one-dimensional hamiltonian
  systems.
\newblock {\em Phys. Rev. Lett.}, 108:180601, 2012.

\bibitem{Barabasi1995}
A-L Barabási and H~E Stanley.
\newblock {\em Fractal concepts in surface growth}.
\newblock Cambridge university press, 1995.

\bibitem{Das2014a}
Suman~G Das, Abhishek Dhar, Keiji Saito, Christian~B Mendl, and Herbert Spohn.
\newblock Numerical test of hydrodynamic fluctuation theory in the
  {Fermi-Pasta-Ulam} chain.
\newblock {\em Phys. Rev. E}, 90(1):012124, 2014.

\bibitem{DiCintio2015}
Pierfrancesco {Di Cintio}, Roberto Livi, Hugo Bufferand, Guido Ciraolo, Stefano
  Lepri, and Mika~J. Straka.
\newblock Anomalous dynamical scaling in anharmonic chains and plasma models
  with multiparticle collisions.
\newblock {\em Phys. Rev. E}, 92:062108, 2015.

\bibitem{Mendl2013}
Christian~B. Mendl and Herbert Spohn.
\newblock Dynamic correlators of {Fermi-Pasta-Ulam} chains and nonlinear
  fluctuating hydrodynamics.
\newblock {\em Phys. Rev. Lett.}, 111:230601, 2013.

\bibitem{Lepri03}
S~Lepri, R~Livi, and A~Politi.
\newblock Universality of anomalous one-dimensional heat conductivity.
\newblock {\em Phys. Rev. E}, 68(6, Part 2):067102, DEC 2003.

\bibitem{Lee-Dadswell2015}
GR~Lee-Dadswell.
\newblock Universality classes for thermal transport in one-dimensional
  oscillator systems.
\newblock {\em Phys. Rev. E}, 91(3):032102, 2015.

\bibitem{Mejia2001}
C.~Mejia-Monasterio, H.~Larralde, and F.~Leyvraz.
\newblock Coupled normal heat and matter transport in a simple model system.
\newblock {\em Phys. Rev. Lett.}, 86(24):5417--5420, 2001.

\bibitem{Casati2009}
Giulio Casati, Lei Wang, and Tomaz Prosen.
\newblock A one-dimensional hard-point gas and thermoelectric efficiency.
\newblock {\em J. Stat. Mech.: Theory and Experiment}, (03):L03004, 2009.

\bibitem{Benenti2014}
Giuliano Benenti, Giulio Casati, and Carlos Mejia-Monasterio.
\newblock Thermoelectric efficiency in momentum-conserving systems.
\newblock {\em New J. Phys.}, 16(1):015014, 2014.

\bibitem{Iubini2016}
S~Iubini, S~Lepri, R~Livi, and A~Politi.
\newblock Coupled transport in rotor models.
\newblock {\em New J. Phys.}, 18(8):083023, 2016.

\bibitem{Iacobucci2011}
A.~Iacobucci, F.~Legoll, S.~Olla, and G.~Stoltz.
\newblock Negative thermal conductivity of chains of rotors with mechanical
  forcing.
\newblock {\em Phys. Rev. E}, 84(6):061108, 2011.

\bibitem{Toda79}
M~Toda.
\newblock Solitons and heat-conduction.
\newblock {\em Phys. Scr.}, 20(3-4):424--430, 1979.

\bibitem{Zotos02}
X~Zotos.
\newblock Ballistic transport in classical and quantum integrable systems.
\newblock {\em J. Low. Temp. Phys.}, 126(3-4):1185--1194, 2002.

\bibitem{Shastry2010}
B~Sriram Shastry and AP~Young.
\newblock Dynamics of energy transport in a {Toda} ring.
\newblock {\em Phys. Rev. B}, 82(10):104306, 2010.

\bibitem{Mazur1969}
P~Mazur.
\newblock Non-ergodicity of phase functions in certain systems.
\newblock {\em Physica}, 43(4):533--545, 1969.

\bibitem{Theod1999}
Nikos Theodorakopoulos and M~Peyrard.
\newblock Solitons and nondissipative diffusion.
\newblock {\em Physical review letters}, 83(12):2293, 1999.

\bibitem{Spohn2018}
Herbert Spohn.
\newblock Interacting and noninteracting integrable systems.
\newblock {\em Journal of Mathematical Physics}, 59(9):091402, 2018.

\bibitem{Kundu2016}
Aritra Kundu and Abhishek Dhar.
\newblock Equilibrium dynamical correlations in the {Toda} chain and other
  integrable models.
\newblock {\em Phys. Rev. E}, 94:062130, Dec 2016.

\bibitem{DiCintio2018}
Pierfrancesco {Di Cintio}, Stefano Iubini, Stefano Lepri, and Roberto Livi.
\newblock Transport in perturbed classical integrable systems: The pinned
  {Toda} chain.
\newblock {\em Chaos, Solitons \& Fractals}, 117:249--254, 2018.

\bibitem{Dhar2019}
Abhishek Dhar, Aritra Kundu, Joel~L. Lebowitz, and Jasen~A. Scaramazza.
\newblock Transport properties of the classical {Toda} chain: Effect of a
  pinning potential.
\newblock {\em Journal of Statistical Physics}, 175(6):1298--1310, Jun 2019.

\bibitem{Iacobucci2010}
Alessandra Iacobucci, Frederic Legoll, Stefano Olla, and Gabriel Stoltz.
\newblock Thermal conductivity of the {Toda} lattice with conservative noise.
\newblock {\em Journal of Statistical Physics}, 140(2):336--348, Jul 2010.

\bibitem{Chen2014}
Shunda Chen, Jiao Wang, Giulio Casati, and Giuliano Benenti.
\newblock Nonintegrability and the {Fourier} heat conduction law.
\newblock {\em Phys. Rev. E}, 90:032134, Sep 2014.

\bibitem{Campa2014}
Alessandro Campa, Thierry Dauxois, Duccio Fanelli, and Stefano Ruffo.
\newblock {\em Physics of long-range interacting systems}.
\newblock OUP Oxford, 2014.

\bibitem{avila2015length}
Ricardo~R Avila, Emmanuel Pereira, and Daniel~L Teixeira.
\newblock Length dependence of heat conduction in (an) harmonic chains with
  asymmetries or long range interparticle interactions.
\newblock {\em Physica A: Statistical Mechanics and its Applications},
  423:51--60, 2015.

\bibitem{Bagchi2017}
Debarshee Bagchi.
\newblock Thermal transport in the {Fermi-Pasta-Ulam} model with long-range
  interactions.
\newblock {\em Phys. Rev. E}, 95(3):032102, 2017.

\bibitem{Olivares2016}
Carlos Olivares and Celia Anteneodo.
\newblock Role of the range of the interactions in thermal conduction.
\newblock {\em Phys. Rev. E}, 94:042117, 2016.

\bibitem{Iubini2018}
Stefano Iubini, Pierfrancesco {Di Cintio}, Stefano Lepri, Roberto Livi, and
  Lapo Casetti.
\newblock Heat transport in oscillator chains with long-range interactions
  coupled to thermal reservoirs.
\newblock {\em Phys. Rev. E}, 97:032102, Mar 2018.

\bibitem{DiCintio2019}
P~{Di Cintio}, S~Iubini, S~Lepri, and R~Livi.
\newblock Equilibrium time-correlation functions of the long-range interacting
  {Fermi{\textendash}Pasta{\textendash}Ulam} model.
\newblock {\em Journal of Physics A: Mathematical and Theoretical},
  52(27):274001, jun 2019.

\bibitem{Ruelle1999}
David Ruelle.
\newblock {\em Statistical mechanics: Rigorous results}.
\newblock World Scientific, 1999.

\bibitem{Vidmar2016}
Lev Vidmar and Marcos Rigol.
\newblock Generalized {Gibbs} ensemble in integrable lattice models.
\newblock {\em Journal of Statistical Mechanics: Theory and Experiment},
  2016(6):064007, 2016.

\end{thebibliography}

\backmatter

\printindex

\end{document}